 \documentclass[conference]{IEEEtran}

\usepackage{graphicx}
\usepackage[para]{threeparttable}
\usepackage{subfig}
\usepackage{array}
\usepackage{mathtext}
\usepackage{enumitem}
\usepackage{multirow}
\usepackage{amsmath}
\usepackage{xcolor}
\usepackage{algorithmic}
\usepackage{algorithm}
\usepackage{epsfig}
\usepackage{url}
\usepackage{epstopdf}
\usepackage{cite}
\usepackage{setspace}
\usepackage{graphicx,lipsum}
\usepackage{caption}
\usepackage{tikz}
\usepackage{footnote}
\makesavenoteenv{tabular}
\makesavenoteenv{table}
\usepackage{color,graphicx}
\usepackage{hyperref}
\usepackage{amsmath}
\usepackage{comment}
\usepackage{pbox}
\usepackage{epstopdf}
\usepackage{epsfig}
\definecolor{Gray}{gray}{0.9}
\newcolumntype{P}[1]{>{\centering\arraybackslash}p{#1}} % Horizontal  Use P for cenetring
\newcolumntype{M}[1]{>{\centering\arraybackslash}m{#1}} % Vertical Use M for centering
\usepackage{xcolor}

\begin{document}
\bibliographystyle{ieeetr}

% \title{Enhancing Resilience of Disaster Networks through Optimized Routing in the AllJoyn Framework}
% \title{Evaluating the Accuracy of Socio-inspired CALM in Predicting Channel Assignment Performance in Simulated Real-world Mesh Networks}

% \title{Evaluation of Socio-inspired Channel Assignment Performance Prediction Metric CALM in Simulated Real-world Mesh Networks}
%\title{Exploring the Statistical Relationship between Theoretical Interference Estimates and Wireless Mesh Network Capacity}
%\title{Exploring the Statistical Relationship between Interference Estimates and Network Capacity}
% \title{Statistical Analysis of Capacity Interference Relationship in Dense and Ultra-Dense Networks}
\title {Roaming Performance Analysis and Comparison between Wi-Fi and Private Cellular Network}
%\title {CIRNO: Leveraging the Capacity Interference Relationship for Optimization of Dense Networks}

% \author{\IEEEauthorblockN{Srikant Manas Kala\textsuperscript \dag, M Pavan Kumar Reddy\textsuperscript \dag, Ranadheer Musham\textsuperscript \dag, and Bheemarjuna Reddy Tamma\textsuperscript \dag}
% \IEEEauthorblockA{ Indian Institute of Technology Hyderabad, India\textsuperscript \dag\\
% \author{\IEEEauthorblockN{Srikant Manas Kala$^\ast$, M Pavan Kumar Reddy$^\ast$, and Bheemarjuna Reddy Tamma$^\ast$}
% \IEEEauthorblockA{ $^\ast$Indian Institute of Technology Hyderabad, India.  $^\dagger$The University of Chicago, Illinois, USA.\\
% Email: [cs12m1012, cs12b1025@iith.ac.in, tbr]@iith.ac.in, vanlin@uchicago.edu}}
% \author{\IEEEauthorblockN{Srikant Manas Kala$^\dag$, Vanlin Sathya$^*$, Betty Lala$^\P$, and~Bheemarjuna~Reddy~Tamma$^\dag$}

%Email: cs12m1012@iith.ac.in, winston.seah@ecs.vuw.ac.nz, vanlin@uchicago.edu, 3es18314s@s.kyushu-u.ac.jp}}
 
\author{\IEEEauthorblockN{Vanlin Sathya, Aasawaree Deshmukh, Mohit Goyal, and Mehmet Yavuz}
%\author{\IEEEauthorblockN{Srikant Manas Kala$^\dag$, Winston K.G. Seah$^\Phi$, Vanlin Sathya$^*$, Betty Lala$^\P$}
%\IEEEauthorblockA{$^\dag$Indian Institute of Technology Hyderabad, India. $^\Phi$Victoria University of Wellington, New Zealand.\\ $^*$University of Chicago, Illinois, USA.\\
\IEEEauthorblockA{Celona. Inc, Cupertino, California, USA.\\
% Email: cs12m1012@iith.ac.in, vanlin@uchicago.edu, 5es18005k@s.kyushu-u.ac.jp, tbr@iith.ac.in}}
Email: vanlin@celona.io, aasawaree@celona.io, mohit@celona.io, mehmet@celona.io}}
%Email: cs12m1012@iith.ac.in, winston.seah@ecs.vuw.ac.nz, vanlin@uchicago.edu, 3es18314s@s.kyushu-u.ac.jp}}

% \author{\IEEEauthorblockN{Srikant Manas Kala$^\ast$, and Bheemarjuna Reddy Tamma$^\ast$}
% \IEEEauthorblockA{ $^\ast$Indian Institute of Technology Hyderabad, India.\\
%  Email: [cs12m1012, tbr]@iith.ac.in}}
\vspace{-0.3cm}
\maketitle
\begin{abstract}
Private network deployment is gaining momentum in warehouses, retail, automation, health care, and many such use cases to guarantee mission-critical services with less latency. Guaranteeing the delay-sensitive application in Wi-Fi is always challenging due to the nature of unlicensed spectrum. As the device ecosystem keeps growing and expanding, all the current and future devices can support both Wi-Fi and Private Cellular Network (CBRS is the primary spectrum in the US for private network deployment). However, due to the existing infrastructure and huge investment in the dense Wi-Fi network, consumers prefer two deployment models. The first scenario is deploying the private network outdoors and using the existing Wi-Fi indoors. The second scenario is to use the existing Wi-Fi network as a backup for offloading the traffic indoors and parallely utilizes the private network deployment for less latency applications.  Hence, we expect, in both scenarios, a roaming between two technologies \emph{i.e.,} Wi-Fi and Private Cellular Network. In this work, we would like to quantify the roaming performance or service interruption time when the device moves from Wi-Fi to Private Network (CBRS) and vice-versa.

\end{abstract}


%%%%%%%%Journal%%%%%%%%%%
%\renewcommand{\jrnl}[1]{\textcolor{blue}{#1}}
\section{Introduction}
In recent days, the mobile device is capable of supporting both Wi-Fi and Citizens Broadband Radio Service (CBRS)~\cite{gao2020performance} technology. 
The recent Wi-Fi 6E supports the operation of Wi-Fi in the unlicensed 6 GHz band. It is effectively an extension of the existing Wi-Fi 6 (or 802.11ax~\cite{naik2020next}) standard (which operates in the 2.4 GHz \& 5 GHz bands), which is known to improve the overall network performance using technologies such as Orthogonal Frequency Division Multiple Access (OFDMA), Basic Service Set (BSS) coloring, target wait times, etc. Thus, the sole benefit of Wi-Fi 6E is using 6 GHz.  

In the current scenario, the density of 6 GHz APs will not increase as much as when you transitioned from 2.4 GHz to 5 GHz~\cite{sathya2020measurement}; it will still increase due to challenges with propagation. Some Wi-Fi vendors recommend designing for 5 GHz as the propagation model so the coverage patterns will not vary as much for 6 GHz. However, to take advantage of the high data rates supported by 6 GHz, you will need more APs in 6 GHz than it is required in a 5 GHz deployment. This obviously results in higher capital and operational costs. Even if you augment your existing Wi-Fi deployment with 6 GHz capable APs and do not do a rip and replace, you will still need additional cable runs/drops, switch ports, power (more on this below), RF design considerations (channel, power) and finally additional management and troubleshooting. 

Is private cellular to be the panacea of all wireless issues? Definitely not. However, it is a welcomed alternative for some latency-sensitive use cases and critical business applications. If you are thinking about 6 GHz or a Wi-Fi refresh, you should definitely consider private cellular. This hope is not only based on the fact that private cellular solves many of the aforementioned limitations of 6 GHz specifically (and Wi-Fi in general) but also on the fact that many of your mission-critical apps and devices will greatly benefit from a reliable, predictable wireless connection. The private network deployment in the US is based on CBRS spectrum availability of 150 MHz in the range of 3.55 GHz to 3.7 GHz channel capacity. Using the CBRS spectrum, enterprises can deploy their private wireless network independently of the licensed spectrum.

Unlike Wi-Fi, where devices contend for the medium and are prone to interference (unlicensed spectrum~\cite{sathya2020measurement}), private cellular~\cite{gao2020performance} is more prescriptive. The network, not the devices, determines how clients connect and roam – effectively facilitating the contention for each AP and device along with determining the channel and power each AP operates at via a central database (Spectrum Access System (SAS~\cite{liu2019design}) in USA). This not only makes the wireless connection reliable but also provides a better connection experience especially when the device is handed off (roams to) to another AP since the core (network) decides when a device has to be handed off. In contrast, in Wi-Fi, the device decides which AP to connect to, at what signal strength, which AP to roam to, and when. If this was not enough, each Original Equipment Manufacturer (OEM) has their own roaming algorithm.

The private cellular network is the turnkey solution that integrates with your existing LAN - whether it is by requesting IP addresses for devices from your DHCP server, routing traffic based on your network configurations, or easily recognizing applications used by devices, seamlessly integrates into the same LAN which is set up for supporting your Wi-Fi networks. Additionally, strong mutual authentication and end-to-end encryption along the full data path, whether wireless or wired, and with MicroSlicing, network segmentation can also be extended over the air.

In most use cases, private cellular will be the best choice compared to Wi-Fi. But due to the existing infrastructure and huge investment in the dense Wi-Fi network, consumers prefer two deployment models. The first scenario is to deploy the private network~\cite{sathya2022comparative,sathya2023warehouse} \emph{i..e,} CBRS outdoors and use the existing Wi-Fi deployment indoors. The second scenario is to use the existing Wi-Fi network as a backup for offloading the traffic indoors and with the new private network deployment for mission critical applications. As the recent device supports both technologies, consumers want to use both interfaces effectively based on the use-case requirement. This work quantifies the service interruption time when the application moves from Wi-Fi to private cellular network and vice-versa.

\begin{table*}[!htb]
\scriptsize
    \caption*{}
     \begin{minipage}{.5\linewidth}
      \centering
        \caption{WLAN Experiment Parameters}
        \begin{tabular}{|c|c|}
            \hline
\bfseries Parameter &\bfseries Value \\ [0.4ex]
\hline\hline
Number of WLAN APs & 4 \\
\hline
WLAN AP Channels & 1, 6 and 11 and 5GHz channels\\
\hline
WLAN Frequency and Band & 2.4 GHz: 20 MHz, 5GHz: 20 MHz\\
\hline
WLAN AP Bandwidth &  20 MHz \\
\hline
Channel Selection & Centralized S/W Controller \\
\hline
Number of WLAN Clients & 5 \\
\hline
WMM & Enabled \\
\hline
WLAN Client Devices & Pixel, iPhone, Motorola, Samsung Xcover Pro  \\
\hline
Monitoring S/W & Wi-Fi Explorer 3 \\
\hline
        \end{tabular}
        \label{wifi}
    \end{minipage}
    \begin{minipage}{.5\linewidth}
      \centering
        \caption{CBRS Experiment Parameters}
        \begin{tabular}{|c|c|}
           \hline
\bfseries Parameter &\bfseries Value \\ [0.4ex]
\hline\hline
Number of Celona APs & 2 \\
\hline
Number of Bandwidth per AP & 40 MHz (20 + 20) \\
\hline
Operating Band & 48 \\
\hline
Operating Frequency &  3.55 to 3.7 GHz
 \\
\hline
Channel Selection & SAS \\
\hline
Micro Slicing & Enabled \\
\hline
MIMO & 2 x 2 \\
\hline
Client Devices & Pixel, Samsung, iPhone, Motorola \\
\hline
Carrier Aggregation & Enabled  \\
\hline

        \end{tabular}
        \label{5G LAN}
    \end{minipage} 
\end{table*}

\begin{figure}[!htb]
 \centering
 \includegraphics[scale=0.35]{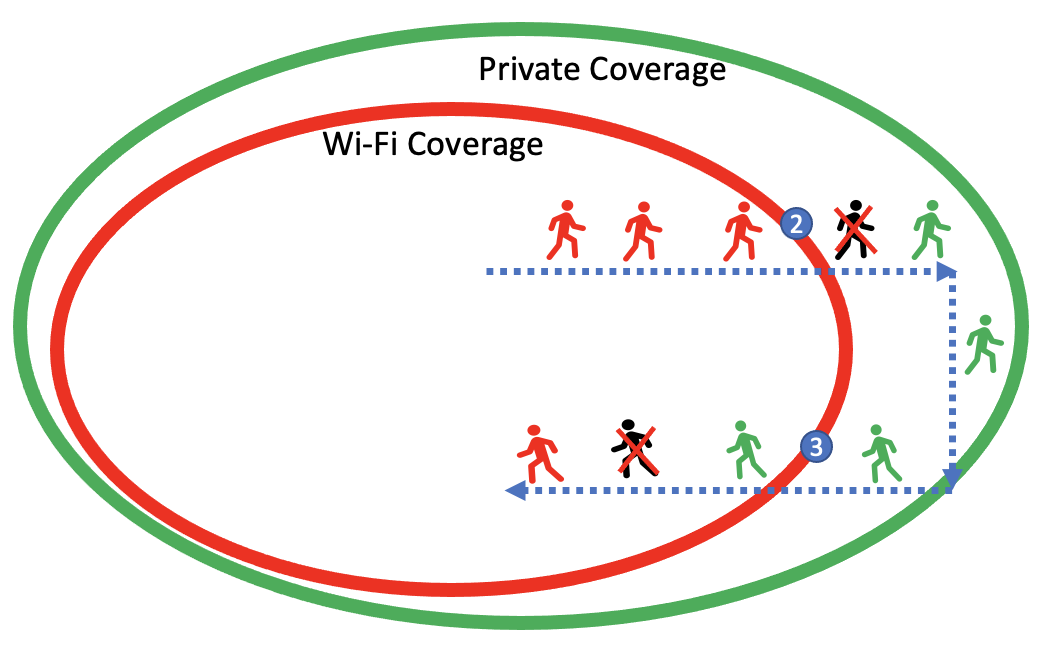}
 \caption{Scenario 1 - Full Private Deployment}
 \label{s11}
\end{figure}

\section{Current Roaming Challenges between Cellular and Wi-Fi}
In the current scenario, the devices prefer Wi-Fi over celluar or private network when Wi-Fi is available. Hence, the cellular to Wi-Fi switching is slow and disrupts the service. In most scenarios, there will not be 100\% private network coverage in all CBRS deployments from day 1. Also, there is no ability to control which applications go over Wi-Fi vs which applications over private networks.
\subsection{Use Case and Scenarios}
\subsubsection{Indoors both Wi-Fi and Private coverage}
In this scenario, we assume the Wi-Fi and private network are deployed in indoor and the device is capable of operating on both technologies.
In the current behavior, the device prefer to use Wi-Fi in indoors and, as it moves outside the coverage of Wi-Fi, sticks to Wi-Fi for a long period (as it is shown in Fig.~\ref{s11} close to the region denoted by number 2) until it switches to the private network and disrupts the service \emph{i.e..,} Wi-Fi $\rightarrow$ private network roaming. Similarly, when the device moves from cellular private network $\rightarrow$ Wi-Fi roaming (which is close to the region denoted by 3), roams quicker compared to Wi-Fi $\rightarrow$ private network. In a real scenario, it is hard to draw a clear boundary between Wi-Fi and private networks, and hence, depending on the radio preference design, it sticks more to that network. The desired behavior is where the device uses a private network for both indoors and outdoors. Hence, the customer cannot disable Wi-Fi because other locations/sites may have Wi-Fi only coverage. It is not practical to switch on/off by the end user.

\subsubsection{Indoors only Wi-Fi and outdoor only CBRS private network coverage}
\begin{figure}[!htb]
 \centering
 \includegraphics[scale=0.35]{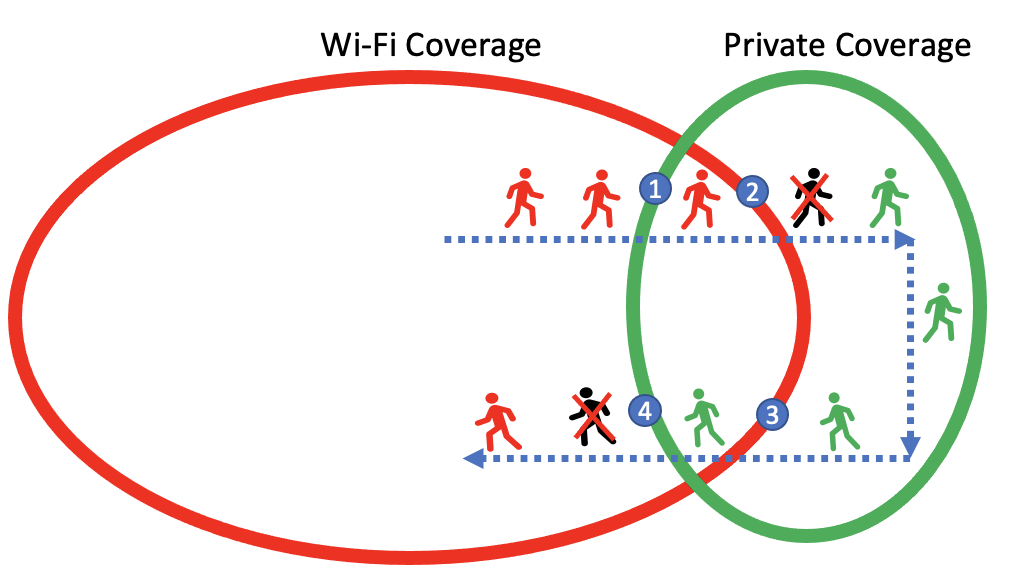}
 \caption{Scenario 2 - Private Coverage Outdoors}
 \label{s22}
\end{figure}

In this scenario, we assume the Wi-Fi is in indoor and private network are deployed in outdoor and the device is capable of operating on both technologies, when it moves from indoor to outdoor and vice-versa.
In the current behavior, the device uses Wi-Fi indoors and, as it moves outside, sticks to Wi-Fi for a long period until it switches to a private network. The disrupts service as shown in Fig.~\ref{s22} \emph{i.e.,} Wi-Fi $\rightarrow$ private network roaming from region 1 to 2. Similarly, the roaming from private network to Wi-Fi roaming \emph{i.e.,} private network $\rightarrow$ Wi-Fi from region 3 to 4. The desired behavior is that the device seamlessly switches to a private network whenever CBRS coverage exists. Compared to Scenario 1, the indoor-only Wi-Fi and outdoor-only private network scenario, the behavior could be better because the Wi-Fi transmission is turned OFF or shut down (either completely or turn OFF specific SSID), so the transition may be quicker.

\begin{figure*}[!htb]

\minipage{0.45\textwidth}
  \includegraphics[width =8cm, height = 4cm ]{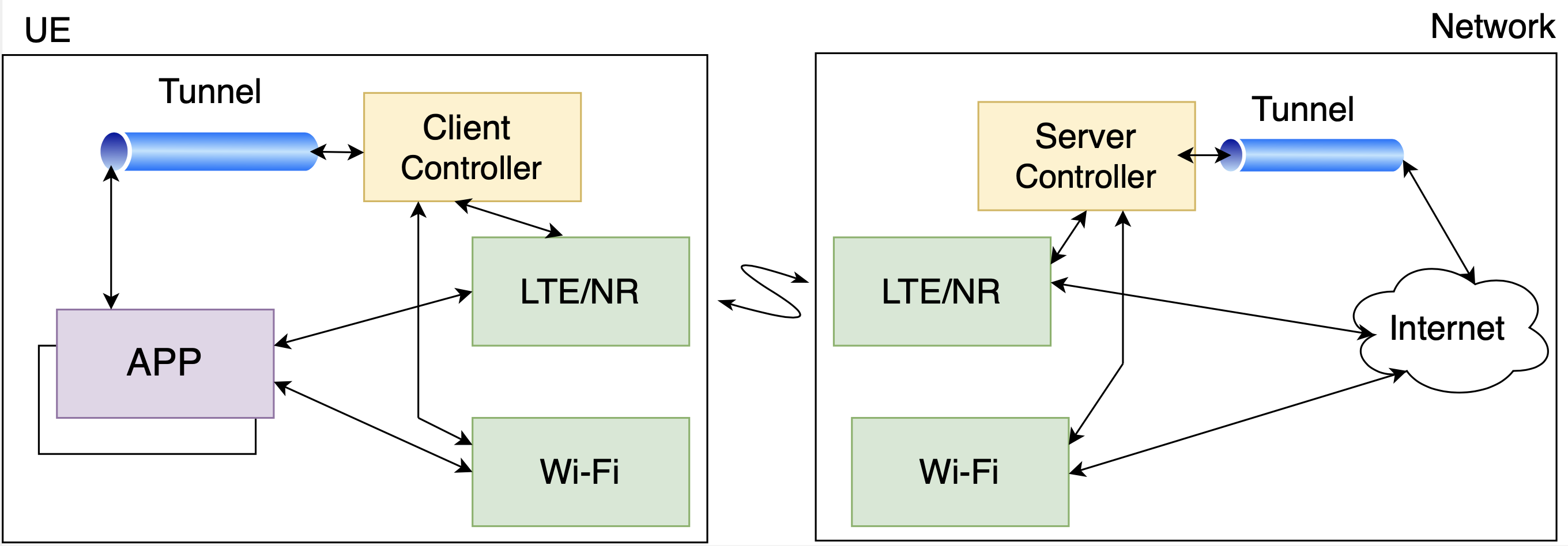}
  \caption*{\textit{(a) Single VPN Tunnel
}}\label{s1}

  \endminipage\hfill
~
\minipage{0.22\textwidth}
  \includegraphics[width =4cm, height = 4cm]{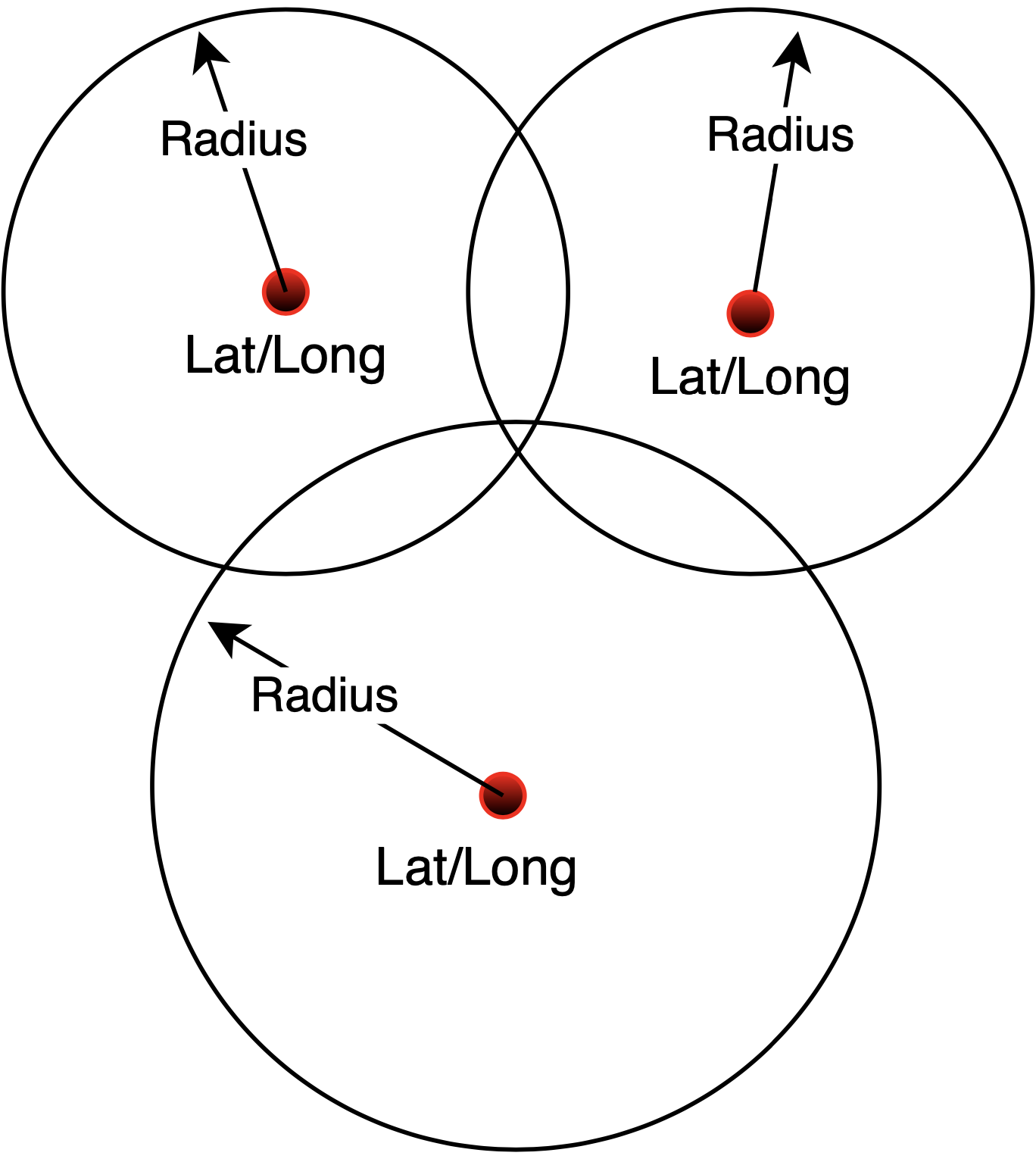}
  \caption*{\textit{(b) Geo-Fencing Boundaries}}\label{s2}

  \endminipage\hfill
  ~
\minipage{0.22\textwidth}
\includegraphics[width =5cm, height = 4.2cm]{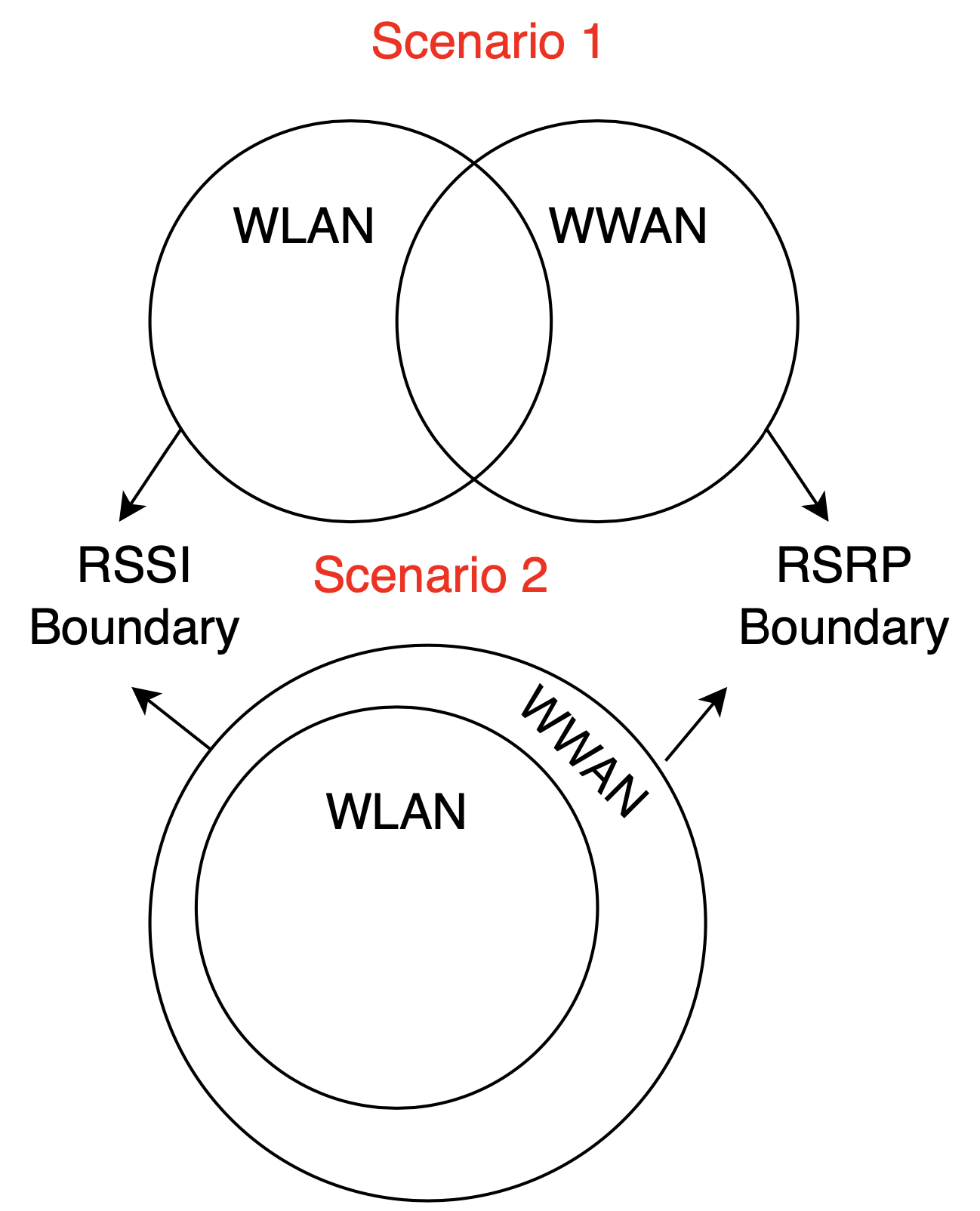}
  \caption*{\textit{(c) RSRP and RSSI Boundaries}}\label{s3}

  \endminipage\hfill

\caption{\textit{Solutions to Improve Roaming Behaviours}}
\label{cs8}
\end{figure*}

\section{Proposed Solutions to Improve the Roaming Behaviours}
\begin{figure}[!htb]
 \centering
 \includegraphics[scale=0.35]{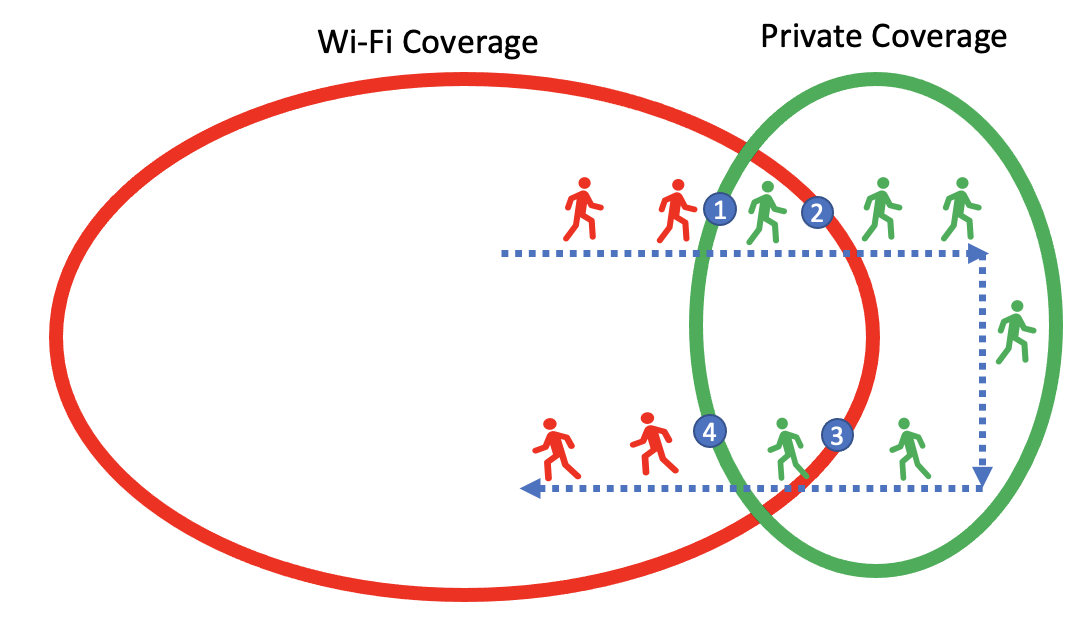}
 \caption{Scenario 4 - Private Coverage Outdoors (Advanced Device Behavior)}
 \label{s4}
\end{figure}

The future mechanism should allow for a seamless transition of traffic across enterprise-LTE or NR and Wi-Fi with both RAT deployed as enterprise solutions. Fig.~\ref{s3} and ~\ref{s4} show the solution design for full private and private coverage outdoor scenarios. Hence, the interruptions shall be less than 1 second. If possible, make-before-break shall be supported. The solution should allow aggregating traffic across enterprise-LTE or NR and Wi-Fi when coverage for both RATs is available. It should allow for prioritizing traffic over enterprise-LTE/NR or Wi-Fi based on
\begin{itemize}
    \item The application preference
    \item Signal strength of LTE or NR and Wi-Fi
    \item Congestion levels on LTE or NR and Wi-Fi networks
    \item Geo-fencing based detection
    \item RSRP and RSSI based detection
\end{itemize}
The solution should allow for specifying the applications that require seamless connectivity and others for which seamless connectivity is optional. It should minimize the compute and packet overheads based on the choice of solutions and selectively enable seamless connectivity based on the application. Furthermore, it allows for enterprise-IT to control the device configuration.

\subsection{Approach 1: Single VPN Tunnel}
The tunnel should be supported over LTE/NR and Wi-Fi IP addresses. This allows continuity with the application layer exposed to a single tunnel inner IP address. Also, it enables dynamic switching RATs based on real-time conditions. There should be enterprise-specific transition policies enabled on top of this framework. This approach needs a client and server, as shown in Fig.~\ref{cs8} (a). The client will run on the device, and the server will run on any centralized controller. The client-side can be a smartphone-based application to support a single tunnel and route the packets over LTE/NR or Wi-Fi based on the RAT availability and preferences for routing decisions. The client application should be capable of detecting the congestion level over LTE/NR and Wi-Fi networks. This can be specified by relative thresholds that require transitioning the traffic across LTE/NR and Wi-Fi.

The server-side application will be running on the enterprise IT to manage the devices from the server app. This can be provided integration with the Mobile Device Management (MDM) platform, so this will be easy to deploy and change the configuration parameters whenever necessary. The server application can route the packets on the Downlink (DL) based on the preferences exercised by the client application for routing the traffic on the Uplink (UL) and reflect that on the DL. The server application shall encapsulate the packet delivered on the DL based on the outer IP address associated with the RAT over which the packet will be sent. Also, the server application can de-encapsulate the packet received on the UL by removing the outer IP address and forwarding the packet to the Intranet or Internet.

\begin{figure}[!htb]
 \centering
 \includegraphics[scale=0.3]{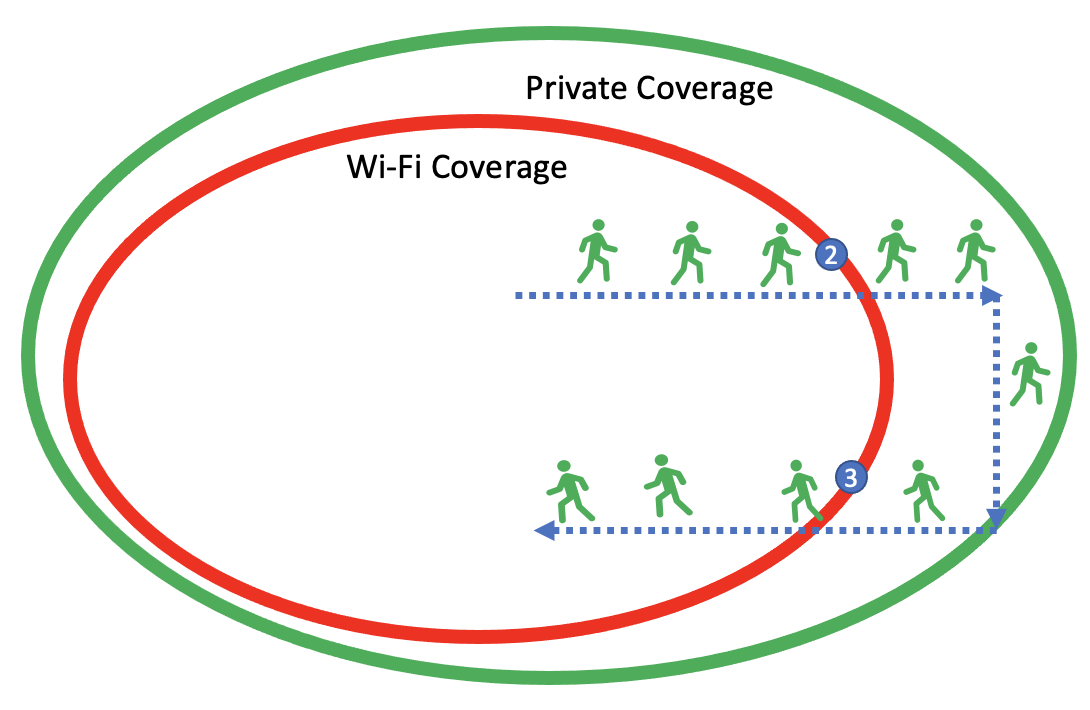}
 \caption{Scenario 3 - Full Private Deployment (Advanced Device Behaviour)}
 \label{s3}
\end{figure}

\subsection{Approach 2: Geo-fencing the Boundaries}
Recently, ONGO Alliance proposed the concept of Geo-fencing private CBRS and non-public networks (such as cellular or Wi-Fi). Employing geofenced areas allows the UEs or devices to perform power-optimized scans when searching for specific Private CBRS / Non-Public network campus network, as shown in Fig.~\ref{cs8} (b). The information provided to the UE includes information on home enterprise networks and can additionally also provide the roaming partner network-related information associated with a given private network subscription. From a network perspective, it primarily allows for better managing common address spaces that may potentially conflict with other deployments.

Private CBRS / Non-Public Networks vary in size and may require coarse geofencing, covering a large area or potentially requiring building-level geofencing to allow the UEs to determine proximity to a Private CBRS / Non-Public Network. Using GPS-based geofencing alone may have power consumption implications on the UE side, and hence, other methods, such as Mobile Network Operator (MNO) network Radio Footprint, may be considered. The GPS location of the Private CBRS / Non-Public network eNB obtained during deployment allows for determining the rough coverage of the campuses. However, it does not directly translate to the actual available RF footprint of the Private CBRS / Non-Public Network.

The geofencing information carried as part of the Enterprise Information is specified as GPS information and/or Radio Footprint information.

\begin{itemize}
    \item GPS information: The geofence information is specified based on the shapes using one or more of these entries to define the boundaries of the Private CBRS / Non-Public network campus.
    \item Radio Footprint information: Specified as a set of public network Cell-IDs indicating a potential availability of a Private CBRS / Non-Public network campus network when the UE enters these macro cells. Also, it is specified as a set of public network Cell-IDs along with the associated parameters (such as SINR, CQI, etc.) to allow for finer control on the locations where the UE should start looking for Private CBRS / Non-Public network campus.
    
\end{itemize}

\subsection{Approach 3: Radio Preferred Configuration}
The network connection manager allows the administrator to set device connection preferences for cellular (WWAN) and Wi-Fi (WLAN) networks, based on RSRP and RSSI, to provide an optimal network connection all the time (as shown in Fig.~\ref{cs8} (c)). It allows administrators to manage and deploy lists of access points for wireless networks, rank them in order of priority, and/or set minimum signal levels when establishing connections, including a set time interval for scanning for Wi-Fi connections. This allows the customer to control the priority of the access technology (Wi-Fi or Cellular), control the priority of the cellular providers, and set a minimum signal level that must be exceeded for the device to attach and use the access technology. This will primarily avoid issues when the Wi-Fi signal is marginal, and the cellular signal will provide better performance.

The device profile is configured in the form of a .txt file, which contains the prioritized list for WWAN and WLAN. The first name on the list is given the highest priority for connection, and each subsequent name will get a lesser priority. Once the configuration is stored on the device, it cannot be edited. A new list file must be created and pushed to the device to add, delete, or change list entries. Also, besides the text file, there are other options to configure the WWAN range, and the allowable RSRP values are in the range of -156 to -31 dBm. Similarly, the WLAN signal level is configured in the range of -90 to -30 dBm. Ultimately, this profile or device configuration is generated as a bar code to be easily loaded on all devices.

 \begin{figure*}[!htb]

\minipage{0.5\textwidth}
  \includegraphics[width =8cm, height = 3.5cm ]{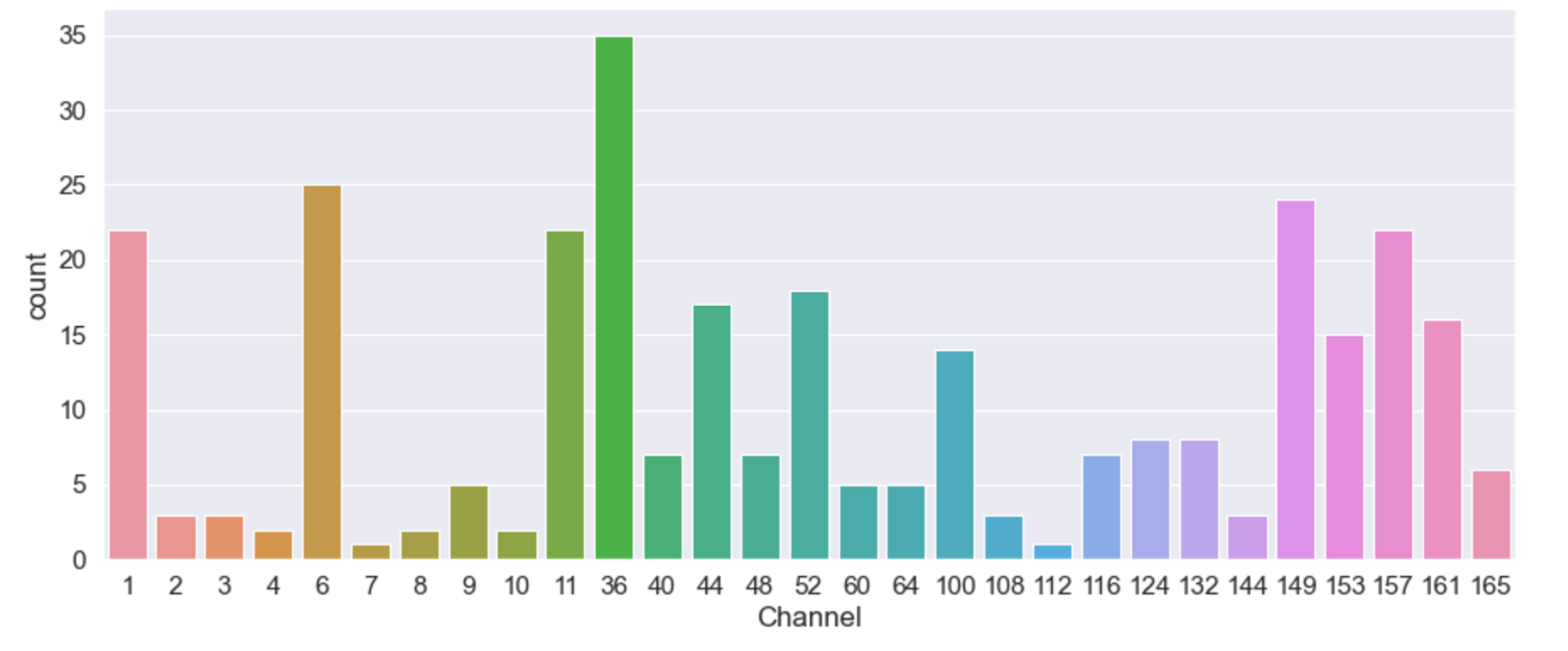}
  \caption*{\textit{(a) Wi-Fi APs on 5GHz Spectrum}}\label{w5}

  \endminipage\hfill
~
\minipage{0.5\textwidth}
  \includegraphics[width =8cm, height = 3.5cm]{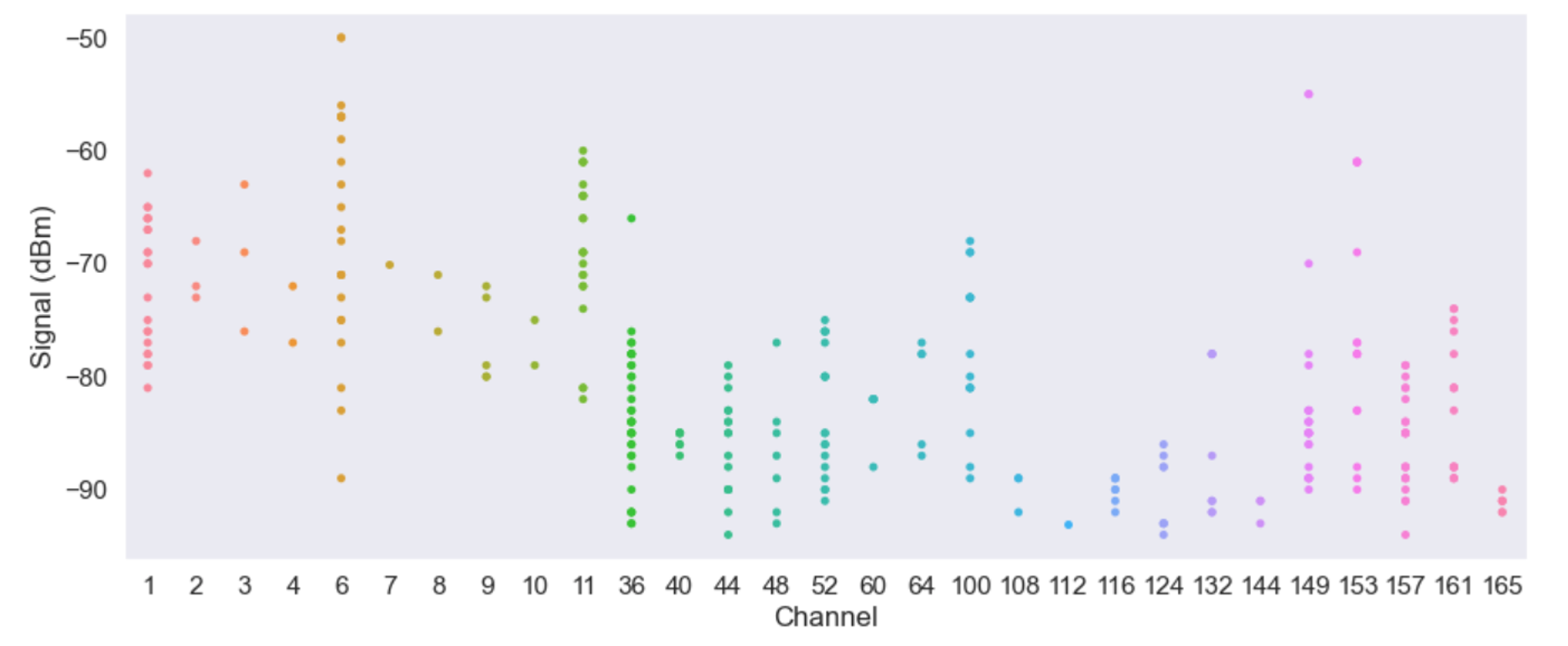}
  \caption*{\textit{(b) Wi-Fi RSSI Signal Levels on 5 GHz Spectrum}}\label{w15}

  \endminipage\hfill

\caption{\textit{Wi-Fi Deployment Survey at Celona HQ}}
\label{ljp}
\end{figure*}

\section{Experiment Environment and Configuration}
This section discusses the real deployment scenario, where Enterprise has deployed Wi-Fi APs as shown in Fig.~\ref{hq1}(a) with a small green circle. Additionally, we deployed a private network in the Celona HQ, as shown in Fig.~\ref{hq1}(a) with a small red circle. Hence, we try to quantify the performance of the two different networks during the roaming scenario.

\subsection{Wi-Fi Environment, Configurations and System Utilization}
Fig.~\ref{hq1} (b) shows the experiment trial area where the roaming experiment is performed. In this setup, we have access to the Wi-Fi network controller because Celona HQ manages the Wi-Fi APs. The tools used to understand the Wi-Fi's Radio Frequency (RF) footprint are Wi-Fi Explorer Pro and Wireshark. We walked at a relatively constant speed outdoors with a mobile cart to understand the boundary region of Wi-Fi coverage in the parking lot area. Fig.~\ref{ljp} (a) shows the number of Wi-Fi APs deployed in the 5 GHz channel. The 5 GHz channels are very crowded by nearby other Wi-Fi APs operating on the same channels. Fig.~\ref{ljp} (b) shows the average Received Signal Strength Indicator (RSSI) observed on 5 GHz Wi-Fi APs. We noticed that all the Wi-Fi 5 GHz operate on all UNII bands with the supporting modes of a, n, ac, and ax. All the Wi-Fi 5 GHz APs are configured to the IEEE Wi-Fi 5 standards, with 20 MHz bandwidth~\cite{sathya2023warehouse}\footnote{In this setup, we try to mimic the warehouse, and supermarket scenarios, where there are dense Wi-Fi APs deployed on 20 MHz configuration in 5 GHz spectrum.}. The maximum number of Wi-Fi streams is 4; therefore the basic rate supported are 6.5, 12, and 24 Mbps. The beacon interval is 102.4 ms, and the average beacon air-time is 0.256 ms. Table~\ref{wifi} shows the Wi-Fi deployment and experiment configuration parameters.
The SSID of the Wi-Fi APs used in the experiments is “Celona” and “Celona-Guest”.

\begin{figure*}[!htb]

\minipage{0.31\textwidth}
  \includegraphics[width =4cm, height = 5cm ]{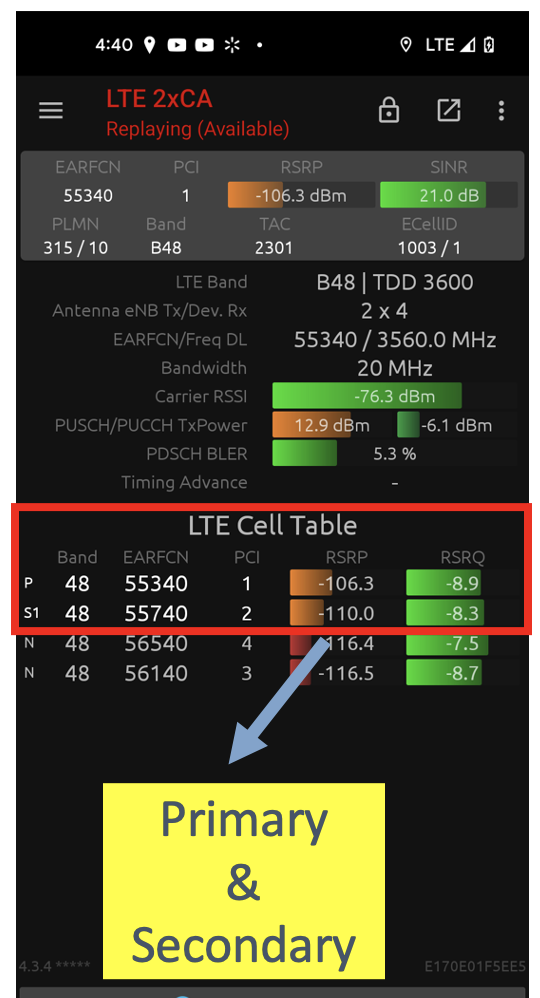}
  \caption*{\textit{(a) Primary and Secondary 
}}\label{lat}

  \endminipage\hfill
~
\minipage{0.31\textwidth}
  \includegraphics[width =4cm, height = 5cm]{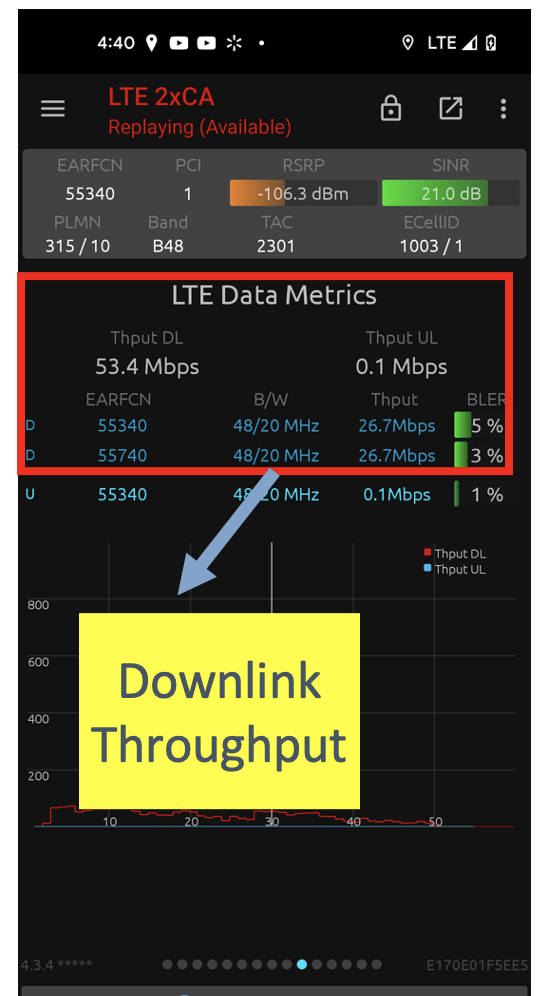}
  \caption*{\textit{(b) CBRS Metrics}}\label{jit}

  \endminipage\hfill
  ~
\minipage{0.31\textwidth}
\includegraphics[width =4cm, height = 5cm]{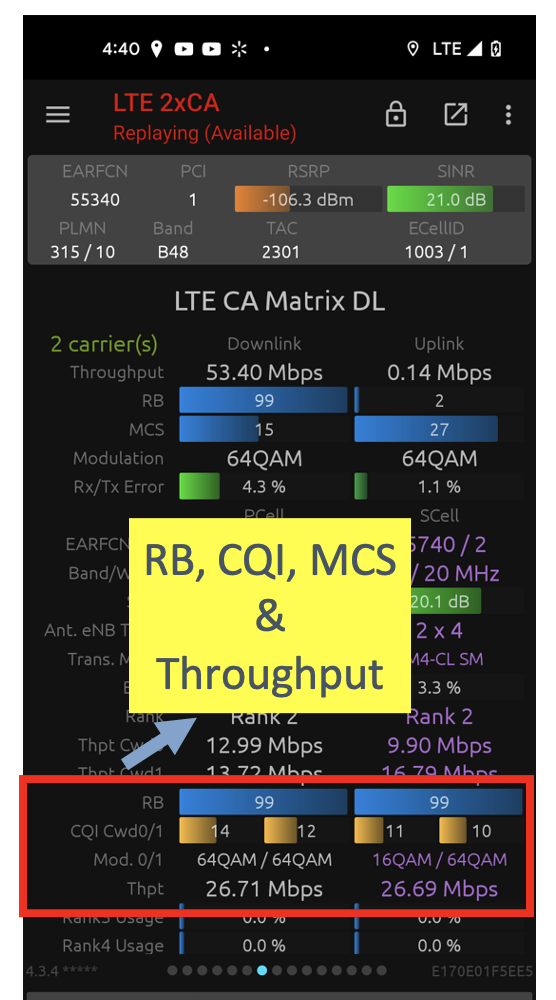}
  \caption*{\textit{(c) RB Allocation}}\label{pd}

  \endminipage\hfill

\caption{\textit{CBRS Survey at HQ Parking Lot }}
\label{cbrs}
\end{figure*}

\subsection{CBRS or Private Network Environment and Configuration}
\begin{figure*}[!htb]

\minipage{0.31\textwidth}
  \includegraphics[width =6cm, height = 4cm ]{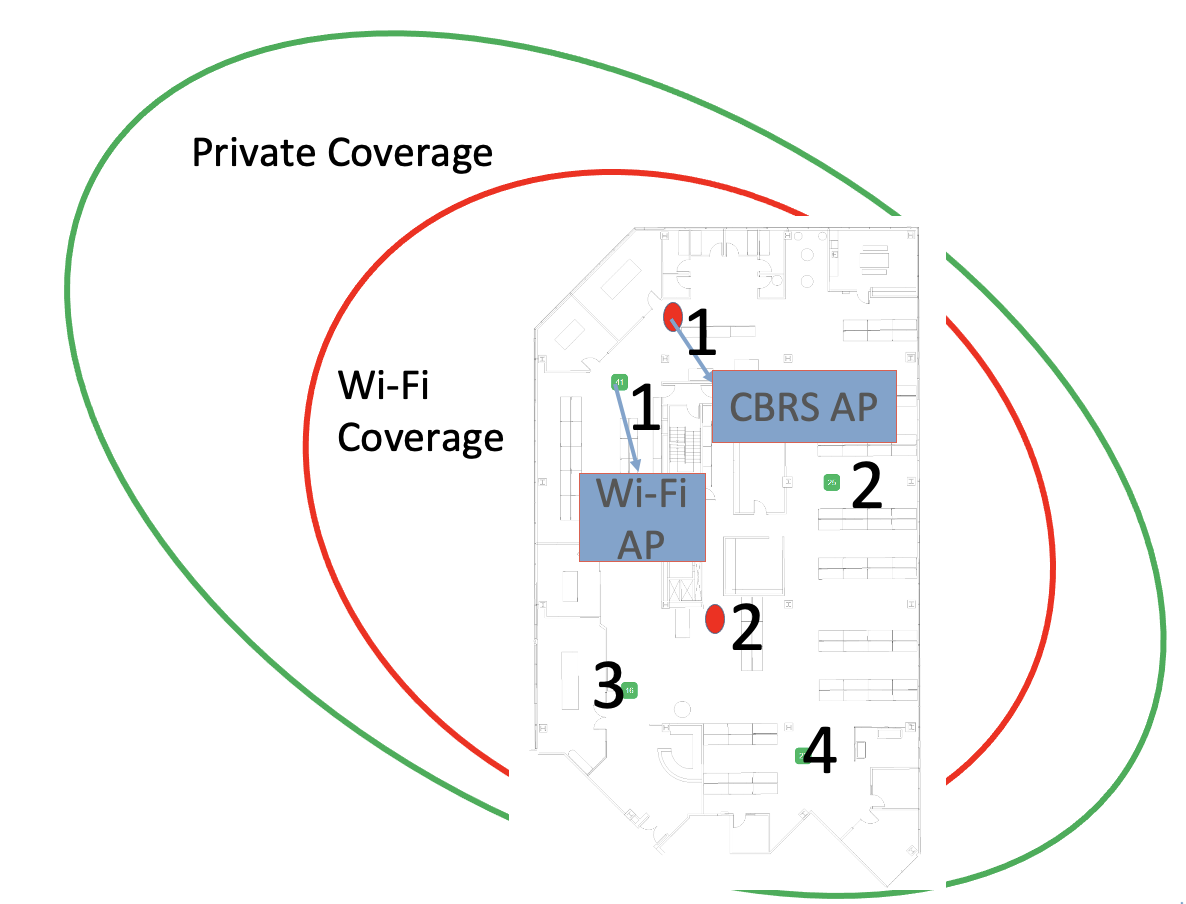}
  \caption*{\textit{(a) Floor Plan}}\label{lat}

  \endminipage\hfill
~
\minipage{0.31\textwidth}
  \includegraphics[width =6cm, height = 4cm]{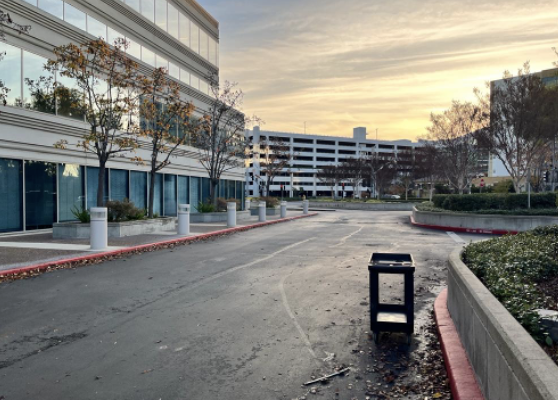}
  \caption*{\textit{(b) Experiment Location}}\label{jit}

  \endminipage\hfill
  ~
\minipage{0.31\textwidth}
\includegraphics[width =6cm, height = 4cm]{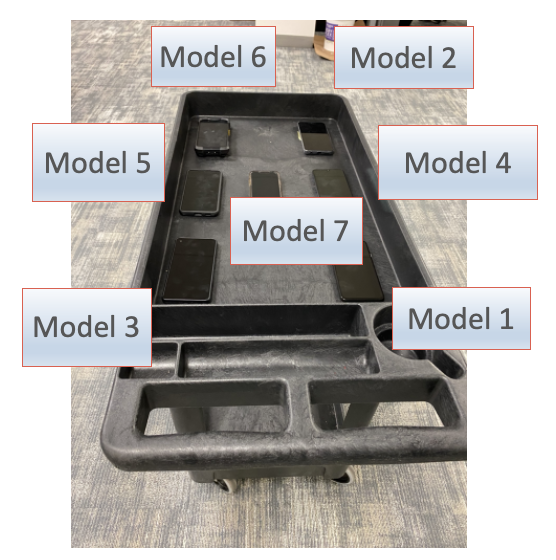}
  \caption*{\textit{(c) Mobile Devices}}\label{pd}

  \endminipage\hfill

\caption{\textit{Celona HQ Experiment}}
\label{hq1}
\end{figure*}

The CBRS deployment (on a private network) and the connection setup, are shown in Fig.~\ref{hq1}(a). Celona Self Organizing Network (SON) algorithm~\cite{5GLAN} is responsible for the frequency planning and setting the optimal EARFCN and transmission power to reduce the co-channel interference. The Physical Cell ID (PCI) allocation algorithm allocates different PCI for each CBRS AP, so there will not be any PCI collision or confusion problem. In this test setup, we used a Network Signal Guru mobile network scanner to collect the radio signals regarding PCI, EARFCN, RSRP, RSRQ, and SINR. The Samsung Galaxy S21+ device is used with the Qualipoc setting to collect all PHY, MAC, and Application layer information. We performed a walk test similar to the Wi-Fi experiment to understand the boundary of the private network in the parking lot region. The collected radio metrics were analyzed using industry-standard post-processing software. Table.~\ref{5G LAN}
shows the number of CBRS AP nodes and other configuration parameters used in the roaming scenario. The Celona APs are operated on 20 + 20 MHz, similar to real use-case deployment scenarios like warehouse, retail, automation, and healthcare. Fig.~\ref{cbrs} (a) shows the primary and secondary carrier of B48 channels, Fig.~\ref{cbrs} (b) shows the throughput shared between the primary and secondary carrier, and Fig.~\ref{cbrs} (c) shows the resource block allocation by both primary and secondary channel \emph{i.e.,} carrier aggregation.

\section{Experimental Setup for Battery Performance}
In this section, we elaborately discuss the experimental test setup of Scenario 1 (as described in Fig.~\ref{s11}), application and traffic scenarios, test procedure, device capabilities, and test cases in terms of execution.

\subsection{Experiment Test Setup}
In this setup, the mobility tests were run in the parking lot of the Celona HQ area, with limited Wi-Fi coverage in a parking lot from indoor Wi-Fi APs leaking outside. On the other hand, good private CBRS coverage in the parking lot from indoor CBRS APs is leaking outside. The experiment goal is to understand the service interruption (\emph{i.e.,} switch time) during roaming between private and Wi-Fi network.

In this experiment setup, we used mobile devices such as Samsung, Google, iPhone, Motorola, and Zebra. In all devices, both cellular and Wi-Fi interfaces are enabled. The Wi-Fi interface is enabled, and the device connects to the Celona HQ's Wi-Fi\footnote{The deployed Wi-Fi AP at HQ supports 802.11 ax with standard 6} network.  The cellular radio interface is connected to the private CBRS network using Celona SIM. From the walking survey in the parking lot area, we observed that the Wi-Fi convergence is shorter than the private network. Hence, we found a region where the Wi-Fi connection is completely down or zero, so we started the experiment from Wi-Fi to the private network region and vice-versa.
\begin{itemize}
    \item \textbf{Continuous Zoom traffic tests:
} In this experiment setup, we run the continuous Zoom traffic session with audio and video turned on, creating an environment where the device/UE continuously uses wireless transmission. All these experiments were conducted at the identified overlap region between the Wi-Fi and private network. 
\begin{figure}[!htb]
 \centering
 \includegraphics[scale=0.38]{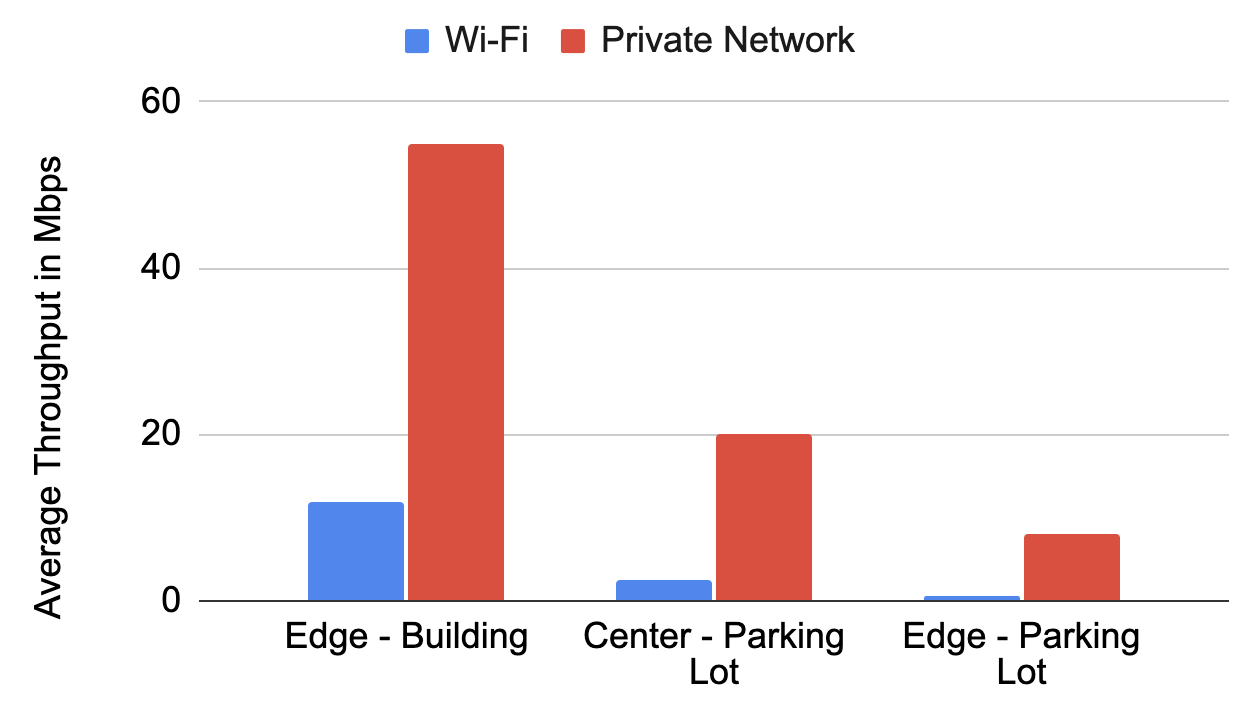}
 \caption{Throughput Performance }
 \label{thp}
\end{figure}

\item \textbf{Shopping Application:}
In this experiment setup, we use the real-time shopping application to understand the behavior of loading the images, adding the items to the cart, and visiting the payment pages during the transition region between Wi-Fi and private network.

\item \textbf{Buffered Traffic Application:}
In this setup, we use YouTube streaming, which runs short videos and tries to understand the end-user performance when the device moves from the transition region between Wi-Fi and private network.

\end{itemize}

\begin{figure}[!htb]
 \centering
 \includegraphics[scale=0.36]{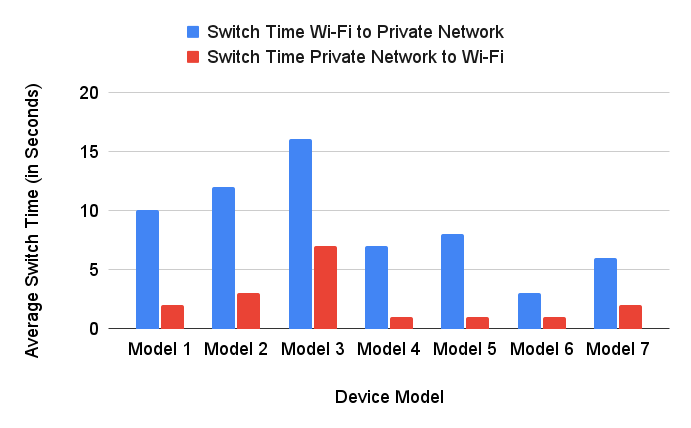}
 \caption{Zoom Live Traffic Results }
 \label{zoom}
\end{figure}

\subsection{Mobile Device Capability}
In this experiment setup, we used different mobile device models primarily used in the USA. Each device has a different model, chipset, Wi-Fi stream, Wi-Fi operating mode, and LTE or NR modes. We observed that Model 6\footnote{Due to privacy limitations and confidential reasons, we are not exposing the device or model name with respect to the roaming behaviour.} devices are expensive than other model devices due to the nature of usage in the warehouse and factory environments. The Model 6 device that we used in this experiment is hard-use tablet, which has a Qualcomm 6490 octa-core, with 2x2 MU-MIMO of Wi-Fi 6E 802.11ax and the LTE and NR B48/N48 support of 4x4 stream parallel transmission. We observed that only Model 3 supports the minimum configurations of a 2x2 cellular transmission stream. 

\subsection{Throughput Performance}
This section observes the throughput performance behavior in the parking lot region for Wi-Fi and private network. Fig.~\ref{thp} shows the average throughput performance on Wi-Fi and private network for three scenarios: Edge to the building, Center region of the parking lot, and edge region of the parking lot. We observed different edge regions for Wi-Fi and private network depending on the transmission power and technologies. Eventually, we observed large coverage leakage from private network due to higher transmission power compared to Wi-Fi. At the edge of Wi-Fi, we observed 500 Kbps from Wi-Fi and 8 Mbps in private network.

 \begin{table}[ht]
 \scriptsize
  \begin{center}
\caption{Shopping Application Performance}
  \begin{tabular}{|p{2.2cm}|p{2.2cm}|p{2.2cm}|}
\hline
 \textbf{Device Model} & \textbf{Switch Time from Wi-Fi to PLTE} & \textbf{Switch Time from PLTE to Wi-Fi
} \\
\hline\hline
Model 1 & 1 S & Seamless \\
\hline
Model 2 & 1 S & 1 S \\
\hline
Model 3 & 3 S & 2 S\\
\hline
Model 4 & 1 S & Seamless\\
\hline
Model 5 & 1 S & Seamless\\
\hline
Model 6 & Seamless & Seamless\\
\hline
Model 7 & 1 S & Seamless \\
\hline
\end{tabular}
\label{app}
\end{center}
\vspace{-0.2cm}
\end{table}

\subsection{Test Environment and Procedure Details}
In the parking lot region, the edge of the Wi-Fi AP RSSI signal level is in the range of -88 to -90 dBm. The CBRS AP coverage of RSRP ranges from -98 to -110 dBm.  All the tests are conducted in the identified boundary region where the roaming transition occurs between Wi-Fi and private network. During the experiment, we moved each device to the boundary and observed the service interruption time.

\subsubsection{Zoom Traffic}
In this section, all devices perform live zoom traffic with audio and video transmission. Fig.~\ref{zoom} shows different mobile device models' average switch time in seconds. The service interruption time or switch time is observed in two ways: (a) switch time from Wi-Fi to private network and (b) switch time from private network to Wi-Fi. In this experiment, we observed that Wi-Fi $\rightarrow$ private network roaming can cause service interruption from 1 second up to 16 seconds. This is mainly because most of the existing devices preferred Wi-Fi\footnote{To avoid unnecessary ping-pong effect from Wi-Fi to cellular and also device vendors prefer to stay in Wi-Fi as long as possible – assuming it is free and cellular incurs a data plan.} so the device try to latch on Wi-Fi network as much as possible (even though it sees better signal on private network). Also, we need to consider that it is not a handover scenario, as it happens in the same RAT when other target AP is better than the serving AP. So, it breaks and makes roaming or handover between the two different RAT technologies. Also, the Wi-Fi $\rightarrow$ private network interruption is significantly higher than private $\rightarrow$ Wi-Fi in service interruption. Also, we observed that the Model 3 device is observed to have the worst performance. This is mainly due to the different chipset vendors. Model 6 and Model 7, devices are observed to have the best performance.

\subsubsection{Shopping Application}
Table~\ref{app} shows the different models and their corresponding switch time from Wi-Fi to private network and vice-versa. We continuously used the shopping application to run the tests by clicking the image every second. We observed that the Wi-Fi $\rightarrow$ private network roaming can cause service interruption for 1 second to 3 seconds. The service interruption is higher when the device roams from Wi-Fi $\rightarrow$ private network compared to the interruption when it roams from private network $\rightarrow$ Wi-Fi. The Model 6 device was observed to have the best seamless user experience. The best part we noticed during the experiment is that the application did not log out, so the device can still maintain the session on both Wi-Fi and private network. However, in both Zoom and shopping applications, the Wi-Fi is usually “sticky,” and the device stays on Wi-Fi as long as possible.

\begin{figure}[!htb]
 \centering
 \includegraphics[scale=0.28]{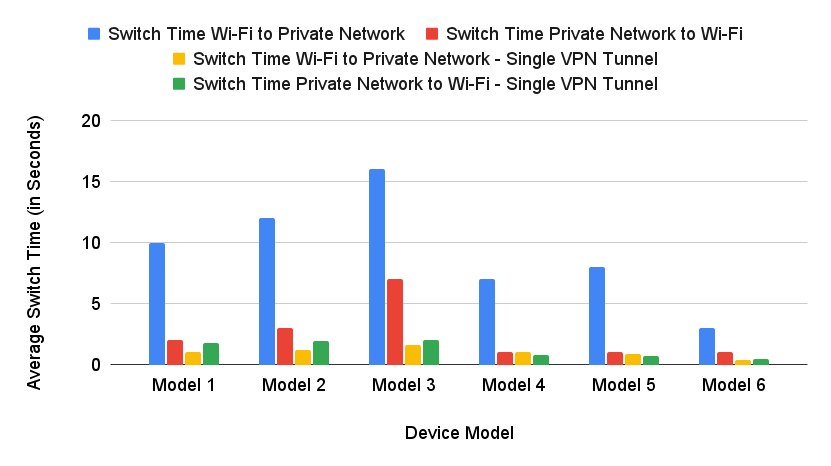}
 \caption{Zoom Live Traffic Results with CCA}
 \label{st1}
\end{figure}

\subsubsection{Buffered Traffic}
In this experiment setup, we observed that if there is YouTube traffic, the user will not feel service interruption from Wi-Fi $\rightarrow$ private network or Vice Versa due to buffer storage. This makes the user feel the continuous streaming even during roaming.

\subsection{Test Results on Single VPN Tunnel}
In this section, we test the single VPN tunnel solution, where the tunnel is supported over LTE/NR and Wi-Fi IP addresses. This eventually allows for continuity with the application layer exposed to a single tunnel inner IP address. Also, it helps the dynamic shown in Fig.~\ref{cs8} (a). We deployed the server app on the core or edge private network, and the client app works on the device side. Each UE enabled with the client app must assign the corresponding server IP address. The private network over the Wi-Fi network particularly prefers this application. Also, for non-critical applications such as streaming \emph{i.e.,} YouTube, packets are unnecessary to send to Tunnel or controller. The reason for not having Model 7 device on the single VPN tunnel result discussion is due to the lack of operating system implementation support on this model.

\subsubsection{Zoom Traffic and Shopping Applications}
In this section, we try to show the benefits of the single VPN-based tunnel solution. Fig.~\ref{st1} shows the average switch time in seconds over different device modes. Here, we try to show four different modes of comparison: (a) Traditional switch time from Wi-Fi $\rightarrow$ private network, (b) Traditional switch time from a private network $\rightarrow$ Wi-Fi, (c) VPN Tunnel switch time from Wi-Fi $\rightarrow$ private network (d) VPN Tunnel switch time from a private network $\rightarrow$ Wi-Fi. We observed that the Wi-Fi $\rightarrow$ private network with single VPN tunnel roaming causes minimum service interruption of 1s to 2s. The Model 6 device is observed to have the best performance. We noticed that the shopping application works seamlessly during roaming time.

\section{Conclusion}
In this work, we analyzed and compared roaming performance between Wi-Fi and private network. After the implementation of single VPN tunnel approach, for the zoom traffic scenario, when the UE roams from Wi-Fi to a private network for the best-case scenario, the service interruption time is less than 1s. Similarly, the service interruption time for the worst-case scenario is from 1s to 2s. With respect to streaming, if there is YouTube traffic, the user won’t feel service interruption from Wi-Fi $\rightarrow$ private network or vice versa due to buffer storage. Also, the shopping application works seamlessly during roaming time. Overall, a single VPN tunnel, is an optimized method to prefer a private network over Wi-Fi, and the UEs stay on a private network as long as possible. Therefore, the overall performance and user experience are seamless without interruption.

\bibliography{ref}

\begin{thebibliography}{1}

\bibitem{gao2020performance}
W.~Gao and A.~Sahoo, ``{Performance Impact of Coexistence Groups in a GAA-GAA
  Coexistence Scheme in the CBRS Band},'' {\em IEEE TCCN}, 2020.

\bibitem{naik2020next}
G.~Naik, J.-M. Park, J.~Ashdown, and W.~Lehr, ``Next generation wi-fi and 5g
  nr-u in the 6 ghz bands: Opportunities and challenges,'' {\em IEEE Access},
  vol.~8, pp.~153027--153056, 2020.

\bibitem{sathya2020measurement}
V.~Sathya, M.~I. Rochman, and M.~Ghosh, ``{Measurement-based coexistence
  studies of LAA \& Wi-Fi deployments in Chicago},'' {\em IEEE Wireless
  Communications}, vol.~28, no.~1, pp.~136--143, 2020.

\bibitem{liu2019design}
S.-Y. Liu, H.-S. Lin, and C.-Y. Huang, ``Design and implement domain proxy
  based cbrs system for 5g,'' in {\em 2019 IEEE VTS Asia Pacific Wireless
  Communications Symposium (APWCS)}, pp.~1--6, IEEE, 2019.

\bibitem{sathya2022comparative}
V.~Sathya, L.~Zhang, and M.~Yavuz, ``A comparative measurement study of
  commercial wlan and 5g lan systems,'' in {\em 2022 IEEE 96th Vehicular
  Technology Conference (VTC2022-Fall)}, pp.~1--7, IEEE, 2022.

\bibitem{sathya2023warehouse}
V.~Sathya, L.~Zhang, M.~Goyal, and M.~Yavuz, ``Warehouse deployment: A
  comparative measurement study of commercial wi-fi and cbrs systems,'' in {\em
  2023 International Conference on Computing, Networking and Communications
  (ICNC)}, pp.~242--248, IEEE, 2023.

\bibitem{5GLAN}
Celona, ``{5G LAN},'' 2020,
  \url{https://assets-global.website-files.com/5e3752187aa7cf8ed3ac0109/628662357eaa01851fdfb744\_Celona\%20Whitepaper\%20-\%20Definitive\%20Guide\%20to\%205G\%20LANs.pdf}.

\end{thebibliography}

\end{document}